\documentclass[creativecommons]{eptcs}

\usepackage{graphicx}

\usepackage[english]{babel}
\usepackage{dsfont}
\usepackage{latexsym}
\usepackage{amssymb}
\usepackage{amsfonts}
\usepackage{fancybox}
\usepackage{wasysym}
\usepackage{epsfig}
\usepackage{fancyhdr}
\usepackage{comment}
\usepackage{breakurl}

\usepackage[tight,figbotcap,tabbotcap,]{subfigure}
\usepackage[all]{xy}
\usepackage{floatflt}

\usepackage{mypaper}
\usepackage{cmll}
\usepackage{proof}
\usepackage{esvect}

\input xy
\xyoption{all}

\usepackage{url}

\newcommand{\escmathcom}[3]{ \newcommand{#1}[#2]{\mbox{#3}}}
\escmathcom{\mcal}{1}{$\mathcal{#1}$}
\escmathcom{\msf}{1}{\sf{#1}}
\escmathcom{\lrelll}{1}{\xleftarrow{#1}}

\escmathcom{\dynex}{0}{$\hat{\exists}$}
\escmathcom{\dynall}{0}{$\hat{\forall}$}
\escmathcom{\lnlambda}{0}{$\hat{\lambda}$}
\escmathcom{\lnLambda}{0}{$\hat{\Lambda}$}
\escmathcom{\caret}{0}{\^{}}
\escmathcom{\lsem}{0}{$\llbracket$}
\escmathcom{\rsem}{0}{$\rrbracket$}
\escmathcom{\lnvareps}{0}{$\hat{\varepsilon}$}
\escmathcom{\One}{0}{$\bf{1}$}
\escmathcom{\Kappa}{0}{$\Box$}
\escmathcom{\Alpha}{0}{$\Theta$}
\escmathcom{\upnu}{0}{$\nu$}
\escmathcom{\lequiv}{0}{$\hat{\equiv}$}
\escmathcom{\downmapsto}{0}{$\downharpoonright$}
\escmathcom{\upmapsto}{0}{$\upharpoonright$}
\escmathcom{\plus}{0}{$\&$}
\escmathcom{\defeq}{0}{$\ =_{df} \ $}
\escmathcom{\ineg}{0}{$\neg$}
\escmathcom{\grax}{0}{$\vDash$}
\newcommand{\freshquant}{\reflectbox{$\mathsf{N}$}}

\escmathcom{\redux}{0}{$ \ \longrightarrow \ $}
\escmathcom{\tranred}{0}{$ \ \rightarrow_{tr} \ $}
\escmathcom{\instred}{0}{$ \ \rightarrow_{ir} \ $}
\escmathcom{\evalred}{0}{$ \ \rightarrow_{er} \ $}
\escmathcom{\tranext}{0}{$ \ \rightarrow_{te} \ $}
\escmathcom{\instext}{0}{$ \ \rightarrow_{ie} \ $}
\escmathcom{\somered}{0}{$ \ \twoheadrightarrow_{r} \ $}
\escmathcom{\patred}{0}{$ \ \rightarrow_{ptr} \ $}
\escmathcom{\paired}{0}{$ \ \rightarrow_{pir} \ $}

\escmathcom{\letexp}{3}{$\msf{let} \ {#1}={#2} \ \msf{in} \ {#3}$}
\escmathcom{\letex}{3}{$\msf{let} \ {#1}={#2} \ \msf{in} \ {#3}$}

\newcounter{observation}
\renewcommand{\theobservation}
{\arabic{observation}}

\newenvironment{obsA}
{\refstepcounter{observation}%
 \begin{description}%
 \item[\textbf{Obs.}
       \theobservation]}%
{\end{description}}

\newcounter{prop}
\renewcommand{\theprop}
{\arabic{prop}}

\newenvironment{propA}
{\refstepcounter{prop}%
 \begin{description}%
 \item[\textbf{Prop.}
       \theprop]}%
{\end{description}}

\title{Resource-Bound Quantification for Graph Transformation} 
\author{Paolo Torrini
\institute{University of Leicester}
\email{pt95@mcs.le.ac.uk}
\and
Reiko Heckel
\institute{University of Leicester}
\email{reiko@mcs.le.ac.uk}
}
\def\titlerunning{Resource-bound quantification}
\def\authorrunning{P. Torrini \& R. Heckel}
\begin{document}
\maketitle
\begin{abstract} Graph transformation has been used to model
  concurrent systems in software engineering, as well as in
  biochemistry and life sciences. The application of a transformation
  rule can be characterised algebraically as construction of a
  double-pushout (DPO) diagram in the category of graphs. We show how
  intuitionistic linear logic can be extended with resource-bound
  quantification, allowing for an implicit handling of the DPO
  conditions, and how resource logic can be used to reason about graph
  transformation systems.
\end{abstract}


\sloppy


\section{Introduction}

Graph transformation (GT) combines the idea of graphs, as a universal
modelling paradigm, with a rule-based approach to specify the
evolution of systems. It can be regarded as a generalisation of term
rewriting. Among the several formalisations of GT based on algebraic
methods, the double-pushout approach (DPO) is one of the most
influential \cite{EEPT06}. Intuitionistic linear logic (ILL) has been
applied to the representation of concurrent systems
\cite{CeSc06,abram93,Mil92}, in relationship with Petri nets, multiset
rewriting and process calculi.  This paper reports work on the
embedding of DPO-GT into a variant of quantified intuitionistic linear
logic with proof terms (HILL). The general goal is to build a bridge
between constructive logic and the specification of concurrent systems
based on graph transformation --- with special attention to
model-driven software development.  Representing model-based
specifications of object-oriented programs as proof terms could be
useful for mechanised verification.

Hypergraphs are a generalisation of graphs allowing for edges that
connect more than two nodes (hyperedges).  Term-based algebraic
presentations of DPO-GT usually rely on hypergraphs and hyperedge
replacement \cite{CHARM}.  Intuitively, an hypergraph can be defined
in terms of parallel compositions of components --- where a component
can be either the empty hypergraph, a node, or an edge component
(an hyperedge with attached nodes). A transformation may delete,
create or preserve components.

It can be convenient to represent an hypergraph as a logic formula,
where hyperedges are predicates ranging over nodes, and composition is
represented by a logic operator. There are naming aspects that need to
be addressed in representing transformation. In particular, (1)
renaming is needed in order to reason about models up to isomorphism,
and (2) the representation of transformation rules involves
abstraction from component names. Transformation cannot be represented
directly in terms of either classical or intuitionistic consequence
relation, because of weakening and contraction.  Accounts based on
hyperedge replacement \cite{CHARM} and second-order monadic logic with
higher-order constructors \cite{Courcelle97} rely on extra-logical
notions of transformation.  Substructural logics offer a comparatively
direct way to express composition as multiplicative conjunction
($\otimes$), and transformation in terms of consequence relation, with
associated linear implication ($\multimap$). This is the case with
linear logic \cite{CeSc06,Dixon06a,Clarke07} as well as with
separation logic \cite{DoPl08}.

There are further semantic aspects to be considered.  One is the
double status of nodes. From the point of view of transformation, each
node as graph component is a linear resource.  From the point of view
of the spatial structure, a node represents a connection between edge
components --- therefore it is a name that may occur arbitrarily many
times. Another aspect is the asymmetry between nodes and edges with
respect to deletion. An edge can be deleted without affecting the
nodes, whereas it makes little sense to delete a node without deleting
the edges it is attached to. On the other hand, by default, edge
deletion should not trigger node deletion. There are systems in which
isolated nodes are disregarded, but this is not generally the case
when dealing with hierarchical graphs
\cite{BusattoKK05,DHP02,HirschM99}, especially in case nodes represent
subgraphs.

We focus on the problem of representing at the object level a
constructive notion of renaming, which behaves injectively, unlike
instantiation of quantified variables and substitution of
meta-variables. Here we rely on a representation of names as terms
that refer to locations, relying on the linear aspect of the logic,
and extending the operational approach presented in \cite{ToHeICE}.
Our goal is more specific than that of higher level approaches to
names with binding based on nominal logic \cite{gabbay04,ScSt04}. In
section \ref{section:GTS} we provide a categorical presentation of
typed DPO-GT, independently of syntactical formalisation. In section
\ref{section:HILL} we present a form of linear lambda calculus with
dependent types, extended with a notion of location (with
$\downmapsto$), and a resource bound quantifier $\dynex$ to represents
name hiding. In section \ref{section:Trans} we show how GT systems can be
embedded in HILL.

\subsection{Overview}

By extending ILL with quantification one can hope to deal with
abstraction, and therefore to reason about GT systems in logic terms
up to $\alpha$-renaming. However, this requires coping with the
difference between variables and names. As a simple example, consider
a graph given by an $r$-typed edge $r(x,y)$ that connects two distinct
nodes $x,y$, and a rule that replaces the $r$-typed edge with a
$b$-typed one, i.e.  $r(n_1,n_2)$ with $b(n_1,n_2)$.  In order to
abstract from node names, assuming $Q_1,Q_2$ are quantifiers, we need
to introduce an abstract representation $Q_1 x y.  r(x,y)$ for the
graph. Intuitively, we could choose between (1) $(Q_1 x y.  r(x,y))
\multimap (Q_1 x y.  b(x,y))$ and (2) $Q_2 x y.  r(x,y) \multimap
b(x,y)$ to represent the rule. It is not difficult to see that no
interpretation of $Q_1,Q_2$ in terms of $\exists,\forall$ is
completely satisfactory. $\exists x y.  b(x,y)$ follows from
$b(n_1,n_1)$, and $\forall x y.  r(x,y)$ implies $r(n_1,n_1)$. In
general, neither existential nor universal quantification can prevent
the identification of distinct variables through instantiation with
the same term --- i.e. they do not behave injectively with respect to
multiple instantiation.

Freshness quantification ($\freshquant$), associated to name
restriction in the context of MF-logic \cite{gabbay04,Pitts01}, relies
on a notion of bindable atom to represent names, an account of
substitution in terms of permutation and of $\alpha$-equivalence in
terms of equivariance.  A typing for restriction can be found in
\cite{ScSt04}.  However, with standard quantifiers, as well as with
freshness, one has that $\exists x.  \alpha$, $\forall x. \alpha$,
$\freshquant x.  \alpha$ are logically equivalent to $\alpha$ whenever
$x$ does not occur in $\alpha$ --- we can call this property
$\eta$-congruence.

In this paper, we define a quantifier ($\dynex$) that keeps the
above-mentioned graph-specific aspects into account --- in particular,
it behaves injectively, and it satisfies the algebraic properties of
name restriction except for $\eta$-congruence. $\dynex$ has a
separating character (though in a different sense from the intensional
quantifiers in \cite{pym02}), by implicitly associating each bound
variable to a linear resource. It has a freshness character in
requiring the relationship between witness terms and bound variables
to be one-to-one --- this makes the introduction rules of $\dynex$
essentially invertible, unlike standard existential quantification.

$\dynex$ can be understood operationally by saying that, with its
introduction, given an instance $M::\alpha[D/x]$, all the occurrences
of the non-linear term $D$ (the witness) in the instance become
hidden, and in a sense the witness becomes linear. In
$\lnvareps(D|n).M ::\dynex x.  \alpha$, the witness may still occur in
the term, but it has been exhaustively replaced with a bound variable
in the type, and it has become associated with the linear location
$n$. We rely on a meta-level representation of hiding in terms of
existential quantification, as usually found in dependent type theory.
The difference lies with the exhaustive character (a freshness
condition) and with the injective association to linear resources.  In
this paper we stop short of introducing restriction $\upnu$ at the
object language level. This could be done, by using as interpretation
for $\dynex$ terms such as $\upnu x.  n_D \otimes M[x/D]$.  However,
extending lambda-calculus with restriction involves more than
technicalities --- see \cite{PiSt93,ScSt04}. Here we limit ourselves
to consider hiding, by using terms such as $\lnvareps(D|n).M = n
\otimes D \otimes M$, with $D$ and $n$ both hidden by the type.


Non-linear terms can be contracted --- i.e. two of the same type can
be merged. This can explain multiple occurrences of a term in an
expression, assuming the point of view of linearity as default.
Technically, the approach we use for names consists of associating the
naming term $D$ to a location, in order to prevent contraction for the
free variables in $D$ (the nominal variables), hence for $D$ itself,
thus closing their scope.  Assuming linearity for locations,
$\eta$-equivalence fails on one hand, and on the other the set of
names turns out minimal --- unlike in \cite{ScSt04}, where the name
space is affine.


\section{Hypergraphs and their transformations} \label{section:GTS}

A hypergraph $(V,E,\msf{s})$ consists of a set $V$ of vertices, a set
$E$ of hyperedges and a function $\msf{s}: E \to V^\ast$ assigning
each edge a sequence of vertices in $V$. A morphism of hypergraphs is
a pair of functions $\phi_V: V_1 \to V_2$ and $\phi_E: E_1 \to E_2$
that preserve the assignments of nodes --- that is, $\phi_V^\ast \circ
\msf{s}_1 = \msf{s}_2 \circ \phi_E$.  By fixing a type hypergraph $TG
= (\mcal{V},\mcal{E},\msf{ar})$, we are establishing sets of node
types $\mcal{V}$ and edge types $\mcal{E}$ as well as defining the
arity $\msf{ar}(a)$ of each edge type $a \in \mcal{E}$ as a sequence
of node types. A $TG$-typed hypergraph is a pair $(HG, type)$ of a
hypergraph $HG$ and a morphism $type: HG \to TG$. A $TG$-typed
hypergraph morphism $f: (HG_1, type_1) \to (HG_2, type_2)$ is a
hypergraph morphism $f: HG_1 \to HG_2$ such that $type_2 \circ f =
type_1$.

A \emph{graph transformation rule} is a span of injective hypergraph
morphisms $L \lrel l K \rrel r R$, called a \emph{rule span}.  A
hypergraph transformation system (GTS) $\mcal{G} = \langle TG, P, \pi,
G_0 \rangle$ consists of a type hypergraph $TG$, a set $P$ of rule
names, a function mapping each rule name $p$ to a rule span $\pi(p)$,
and an initial $TG$-typed hypergraph $G_0$.  A \emph{direct
  transformation} $G \Rrel{p, m} H$ is given by a \emph{double-pushout
  (DPO) diagram} as shown below, where (1), (2) are pushouts and top
and bottom are rule
spans. 
For a GTS $\mcal{G} = \langle TG, P, \pi, G_0 \rangle$, a derivation
$G_0 \Rrel{} G_n$ in $\G$ is a sequence of direct transformations $G_0
\Rrel{r_1} G_1 \Rrel{r_2} \cdots \Rrel{r_n} G_n$ using the rules in
$\G$. An hypergraph $G$ is \emph{reachable} in \mcal{G} iff there is a
a derivation of $G$ from $G_0$.


\centerline{ \xymatrix{ L  \ar@{}[dr]|{(1)} \ar@{->}[d]_{m} & K
\ar@{}[dr]|{(2)} \ar[l]_{l} \ar[r]^{r} \ar@{->}[d]^{d} &
R  \ar@{->}[d]^{m^*}\\
G  & D  \ar[l]^{g}_{} \ar[r]^{}_{h} & H}}

Intuitively, the left-hand side $L$ contains the structures that must
be present for an application of the rule, the right-hand side $R$
those that are present afterwards, and the gluing graph $K$ (the
\emph{rule interface}) specifies the ``gluing items'', i.e., the
objects which are read during application, but are not consumed.
Operationally speaking, the transformation is performed in two steps.
First, we delete all the elements in $G$ that are in the image of $L
\setminus l(K)$ leading to the left-hand side pushout (1) and the
intermediate graph $D$.  Then, a copy of $L \setminus l(K)$ is added
to $D$, leading to the derived graph $H$ via the pushout (2).  The
first step (deletion) is only defined if the built-in application
condition, the so-called gluing condition, is satisfied by the match
$m$.  This condition, which characterises the existence of pushout (1)
above, is usually presented in two parts.
\begin{description}
\item [Identification condition:] Elements of $L$ that are meant to be
  deleted are not shared with any other elements --- i.e., for all $x
  \in L \setminus l(K), \ y \in L$, $m(x) = m(y)$ implies $x=y$.
\item [Dangling condition:] Nodes that are to be deleted must not
  be connected to edges in $G$ that are not to be deleted --- i.e.,
  for all $v \in G_V$, for all $e \in G_E$ such that $v$ occurs in
  $\msf{s}(e)$, then $e \in m_E(L_E)$.
\end{description}

The first condition guarantees two intuitively separate properties:
first --- nodes and edges that are deleted by the rule are treated
linearly, i.e., $m$ is injective on $L \setminus l(K)$; second ---
there must not be conflicts between deletion and preservation, i.e.,
$m(L \setminus l(K))$ and $m(l(K)$ are disjoint. The second condition
ensures that after the deletion action, the remaining structure is
still a graph, and therefore does not contain edges short of a node.


As terms are often considered up to renaming of variables, it is
common to abstract from the identity of nodes and hyperedges
considering hypergraphs up to isomorphism. However, in order to be
able to compose graphs by gluing them along common nodes, these have
to be identifiable. Such potential gluing points are therefore kept as
the \emph{interface} of a hypergraph, a set of nodes $I$ (external
nodes) embedded into $HG$ by a morphism $i: I \to HG$.  An abstract
hypergraph $i: I \to [HG]$ is then given by the isomorphism class
$\{i': I \to HG' \mid \exists \mbox{ isomorphism } j: HG \to HG'
\mbox{ such that } j \circ i = i'\}$.

If we restrict ourselves to rules with interfaces that are
\emph{discrete} (i.e., containing only nodes, but no edges), a rule
can be represented as a pair of hypergraphs with a shared interface
$I$, i.e., $\Lambda I. L \Rrel{} R$, such that the set of nodes $I$ is
a subgraph of both $L, R$.  This restriction does not affect
expressivity in describing individual transformations because edges
can be deleted and recreated, but it reduces the level concurrency.
In particular, concurrent transformation steps can no longer share
edges because only items that are preserved can be accessed
concurrently.

Syntactical presentations of GT based on this semantics have been
given, relying on languages with a monoidal operator, a name
restriction operator and an appropriate notion of rule and matching
\cite{CHARM}.

\section{Linear lambda-calculus} \label{section:HILL}
 
We give a constructive presentation of an extension of intuitionistic
linear logic based on sequent calculus, using a labelling of logic
formulas that amounts to a form of linear $\lambda$-calculus
\cite{abram93,BeEA93,CePf02,Pf02}. We build on top of a system with
ILL propositional type constructors $\multimap, \otimes, \One, !$ and
universal quantifier $\forall$ (we omit $\to$ as case of the latter).
Each of these can be associated to a $\lambda$-calculus operator
\cite{abram93,Pf02}.  Linear implication ($\multimap$) is used to type
linear functions, and we use $\lnlambda$ for linear abstraction (with
$\caret$ for linear application), to distinguish it from non-linear
$\lambda$ (typed by $\forall$). We assume $\alpha$-renaming and
$\beta$-congruence for $\lambda$ and $\lnlambda$ (with linearity check
for the latter). The operator associated to $\otimes$ is parallel
composition, with $\msf{nil}$ as identity.  The ! is interpreted as
closure operator. We extend this system with a dependant type
constructor $\downmapsto$ to introduce a notion of naming, and with a
resource-bound existential quantifier $\dynex$ associated with linear
hiding.

We rely on a presentation based on double-entry sequents
\cite{Pf94,Pf02}. A sequent has form $\Gamma; \Delta \vdash
N::\alpha$, where $\Delta$ is the linear context, as list of typed
linear variables ($v,u, \ldots$) among which we distinguish location
variables ($n,m,\ldots$), and $\Gamma$ is the non-linear context, as
list of typed variables ($x,y,\ldots$). We implicitly assume
permutation and associativity for each context, and use a dot
($\cdot$) for the empty one.  $N::\alpha$ is a typing expression
(typed term) where $N$ is a label (term) and $\alpha$ is a logic
formula (type). 
$\vdash$ represents logic consequence, whenever we forget about terms.
We need to keep track of the free nominal variables in order to
constrain context-merging rules, and to this purpose we annotate each
sequent with the set $\Sigma$ of such variables, writing $[\Sigma];
\Gamma; \Delta \vdash N::\alpha$, where $\Sigma$ is a subset of the
variables declared in $\Gamma$.

Derivable sequents are inductively defined from the axioms and proof
rules in section \ref{section:rules}, and with them the sets of
well-formed terms and non-empty types. Notice that the definition of
derivation includes that of the free nominal variable set $\Sigma$.

Syntactically, terms are $ M = v \mid x \mid n \mid \msf{nil} \mid N_1
\otimes N_2 \mid \lnvareps (D|N). N \mid \lambda x. N \mid \lnlambda
u. N \mid N_1 \caret N_2 \mid N D \mid !N \mid \msf{discard} \ \Gamma
\ \msf{in} \ N \mid \msf{copy}(x) $, where non-linear terms (those
that do not contain free linear variables) are $D = x \mid !N$.
Formulas (or types) are $ \alpha = A \mid
E(D_1:\alpha_1,\ldots,D_n:\alpha_n) \mid \One \mid \alpha_1 \otimes
\alpha_2 \mid \alpha_1 \multimap \alpha_2 \mid ! \alpha_1 \mid \forall
x:\beta. \alpha \mid \dynex x:\beta.  \alpha \mid \alpha \downmapsto D
$. Linear equivalence is defined by $\alpha \lequiv \beta \ \defeq
\ (\alpha \multimap \beta) \otimes (\beta \multimap \alpha)$.
Patterns are terms given by $ P = v \mid x \mid n \mid \msf{nil} \mid
P_1 \otimes P_2 \mid \lnvareps (x|n). P \mid !P \mid \msf{copy}(x) $.
They are used in \emph{let} expressions, defined as $\quad \msf{let}
\ P = N_1 \ \msf{in} \ N_2 \ \defeq \ N_2[N_1/P]$.

We say that $\gamma$ is an \emph{atomic type} whenever either $\gamma
= A_i$ or $\gamma = E_i(D_1:\alpha_1,\ldots,D_k:\alpha_k)$ where
$\alpha_1,\ldots,\alpha_k$ are closed types. We take $A_0,A_1,\ldots$
to be atomic closed types, meant to represent GT node types. A node of
type $A$ is represented as non-linear variable of type $!A$ (see
section \ref{section:Trans}).  We take $E_0,E_1,\ldots$ to stand for
atomic type constructors, meant to be associated with GT edge types. A
HILL type $E(D_1:T_1,\ldots,D_k:T_k)$ (by annotating terms with their
types) is meant to represent a GT edge type $E(A_1,\ldots A_k)$, if we
forget node terms, whenever $T_1 = \ !A_1, \ldots T_k = \ !A_k$.

Semantically, we assume that $v \in LV$, a set of linear variables, $x
\in UV$, a set of non-linear variables, and $n \in LOC \subset LV$, a
set of linear variables that evaluate to themselves and that we call
locations. Given a derivable sequent $\Omega$, the non-linear context
$\Gamma$ can be interpreted as a partial function $UV \to TY$ such
that $\Gamma(x)$ is either closed or undefined for each $x$, and the
linear context $\Delta$ as a partial function $LV \to TY$, such that
for each $n \in LOC$, if defined $\Delta(n)$ has form $\alpha
\downmapsto D$ (\emph{location type}) with $\alpha$ closed, and $D$
non-linear term of type $\alpha$. The free variables in $\Omega$ are
those for which either $\Gamma$ or $\Delta$ is defined.
$FV_{\Omega}(N)$ denotes the free variables occurring in $N$,
$FV_{\Omega}(\alpha)$ those occurring in $\alpha$ (subscripts omitted
in case of no ambiguity). We require for $\Delta_{|LOC}$ (restriction
of $\Delta$ to $LOC$) to satisfy the following \emph{separation
  condition}: for each $n, m \in LOC, n \neq m$ if defined
$FV(\Delta_{|LOC}(n)) \cap FV(\Delta_{|LOC}(m)) = \emptyset$.  We say
that a location is \emph{proper} if $FV(\Delta_{|LOC}(n)) \neq
\emptyset$, \emph{improper} otherwise.

The location typing assignment $n:\beta \downmapsto D$ says that $D$
of type $\beta$ is the \emph{naming term} of $n$, that $n$ is the
location ($\beta$-location) of $D$, and that the variables that occur
free in $D$ (\emph{nominal variables}) are located at $n$.  We denote
by $Names_{\Omega}$ the subset of well-typed terms that occur as
naming terms in $\Omega$. We use $FN(D)$ (resp.  $FN(\alpha)$) to
denote the nominal variables that occur free in $D$ (resp.  $\alpha$),
and we denote by $\Sigma$ the set of the free nominal variables in
$\Omega$, i.e. the free variables that occur in $Names_{\Omega}$.
Variables become nominal when located.  Semantically, a name can be
thought of as a pair $(D,n)$ (naming term and location). The
separation condition implies that $\Delta_{LOC}$ is injective in a
strong sense --- different locations are associated with names that do
not share free variables.

The separation condition required by the definition of $\Delta_{|LOC}$
needs to be enforced explicitly, in all the context-merging rules. In
order to express the constraint, we annotate sequents with the
recursively computed set $\Sigma$ (in brackets) of the free naming
variables. We take $[\Sigma,\Sigma',x]$ to represent the disjoint
union $\Sigma \uplus \Sigma' \uplus \{ x \}$, and $[\Sigma - x]$ to
represent $\Sigma \backslash \{ x \}$ if $x \in \Sigma$ and $\Sigma$
otherwise.  The introduction of locations determines a change in the
behaviour of the free non-linear variables that become nominal: by the
separation condition, two free nominal variables with different
locations cannot be identified.  This corresponds to restricting the
application of meta-level contraction --- as implicit in the
double-entry sequent formulation. Rule \emph{Contr} in the proof
system has a more technical character \cite{Pf02} and it is unaffected
by the separation condition.

The rule $\dynex R$ introduces $\downmapsto$ on the left, whereas
$\dynex L$ eliminates it. Notice that $\downmapsto$ is not treated as
standard constructor in the rules --- we do allow it to appear in
positive position with proper locations. There are no axioms and no
right introduction rules for $\downmapsto$, and it is not possible to
derive a proper location from $\Gamma$, as all variables declared in
$\Gamma$ are of closed types. With the given restriction in place,
only improper locations can be un-linearised, i.e.  $\vdash (!  \alpha
\downmapsto D) \multimap \alpha \downmapsto D$ with $D$ closed, and
moreover $\vdash (!\forall x.  \alpha \downmapsto x) \multimap \forall
x.  \alpha \downmapsto x$, but $\nvdash \alpha \downmapsto D \multimap
\alpha \downmapsto D$.

Intuitively, the $\dynex L$ rule binds a name (a naming variable and a
location), extending the schema of the standard existential rule.  The
$\dynex R$ rule creates a name and hides it (both naming term and
location), replacing exhaustively the term with a bound variable in
the type. Notice that locations may occur in negative positions either
free, bound (with $\multimap$) or hidden bound (with $\dynex$), and
may occur in positive positions only hidden (with $\dynex$), whether
bound or free. A term is a \emph{location term} when it evaluates to a
location. As there is no right introduction of $\downmapsto$, we do
not need to consider complex location terms explicitly.  The operator
associated to $\dynex$ can be defined as

$$
\lnvareps (D|n). M :: \dynex x: \beta. \alpha \ \defeq \ (D::\beta)
\otimes (M::\alpha[D/x]) \otimes (n::\beta \downmapsto D)
$$

for a non-linear term $D ::\beta$, with closed $\beta$ and $x$ not occurring
in $D$, that additionally satisfies a \emph{freshness condition}: $FV(D) \cap
FV(\alpha) = \emptyset$.

The definition of $\lnvareps$ is based on that of proof-and-witness pair
associated with the interpretation of existential quantifier, in standard
$\lambda$-calculus \cite{ProofTh00} as well as in its linear version
\cite{CePf02,Pf02} --- however, here a location is added as evidence that the
witness is located.  The location $n$ is a linear term --- this changes the
nature of the operator, giving it a resource-bound character.

The freshness condition ensures that the occurrences of the name are
the same as the occurrences of the naming term in the main type, and
makes the introduction rules of $\dynex$ essentially invertible,
unlike standard existential quantification. The freshness condition is
trivially satisfied in the case of $\dynex L$. In the case of $\dynex
R$, it follows from the fact that $\alpha \to \alpha$ can be derived
from $\Gamma_1,x$, whereas $D$ can be derived from $\Gamma_2$ ---
assuming that $\Gamma_1$ and $\Gamma_2$ are disjoint, and that $x$
does not occur in $D$. Unlike standard linear logic rules, the
definition of $\dynex R$ involves splitting the non-linear context.

The following statements can be proved by induction on the definition
of derivation, using the separating condition and linearity of
locations. Unlike in double-entry formulations of standard ILL, rule
Weakening is explicitly needed here, in order to prove Cut elimination
for the $\dynex$ case.

\begin{propA} \label{obs:51} (1) Rules \emph{Cut} and \emph{!Cut} can
  be eliminated without loss for provability. \\
  (2) Given a derivation $[\Sigma]; \Gamma; \Delta \vdash N::\alpha$ \\
  \quad (2.a) it is possible to define a surjective function $Loc$
  from the free nominal variables in $\Sigma$ to the set of the naming
  terms $Names$, such that $Loc(x) = D$ iff $x \in FN(D)$ and $D \in
  Names$.\\
  \quad (2.b) given a non-linear closed type $\alpha$ such that
  neither a closed term $D::\alpha$ nor a term of type $\forall
  x:\alpha.  \alpha \downmapsto x$ are derivable from $\Gamma$, there
  is a one-to-one correspondence between the $\alpha$-locations in
  negative positions and those (hidden) in positive positions.
\end{propA}

\begin{propA} \label{obs:5} The following formulas are provable \\

$\vdash (\dynex x:\alpha. \beta) \ \lequiv \ (\dynex y:\alpha.
\beta[y/x]) \qquad$ ($y$ not in $\beta$)

$\vdash (\dynex x y:\alpha. \gamma) \ \lequiv \ (\dynex y x:\alpha.
\gamma)$

$\vdash (\dynex x:\alpha. \beta \otimes \gamma) \ \lequiv \ (\beta
\otimes \dynex x:\alpha. \gamma) \qquad$ ($x$ not in $\beta$)

$\vdash (\dynex x:\alpha. \beta \multimap \gamma) \ \multimap \ (\beta
\multimap \dynex x:\alpha. \gamma) \qquad$ ($x$ not in $\beta$)
\end{propA}

Notice that in general, an operator $\upnu$ can be characterised as
name restriction when it satisfies the following properties
\cite{CHARM}.

$\alpha$-renaming: $\qquad \upnu y. N \ \equiv \ \upnu z. N[z/y]$,
avoiding variable capture

permutation: $\qquad \upnu x y. N \ \equiv \ \upnu
y x.  N$

scope extrusion: $\qquad \upnu x. N_1 \otimes N_2 \ \equiv \
N_1 \otimes \upnu x. N_2$, with $x$ not in $N_1$

$\eta$-congruence: $\qquad \upnu x. N \ \equiv \ N$, with $x$
not in $N$

By the first three formulas in Prop. \ref{obs:5}, $\dynex$ satisfies
properties of $\alpha$-renaming, exchange and distribution over
$\otimes$, and therefore $\lnvareps$ satisfies the corresponding
properties of restriction. On the other hand, $\dynex$ does not
generally satisfy $\eta$-congruence, i.e.  it cannot be proved that
$\alpha$ is equivalent to $\dynex x. \ \alpha$ when $x$ does not occur
free in $\alpha$ (neither sense of linear implication holds).

It is not difficult to see that the following formulas, which are all
valid for existential quantification, fail for $\dynex$

\begin{propA} \label{obs:4}
(1) \ $\nvdash (\dynex x:\beta. \
  \alpha(x,x)) \ \multimap \ \dynex x y:\beta. \ \alpha(x,y)$

(2) \ $\nvdash \forall x:\beta. \ (\dynex z:\beta. 
\alpha(z,z)) \ \multimap \ \dynex y:\beta.
\alpha(y,x)$

(3) \ $\nvdash (\dynex y x:\beta. \ \alpha_1(x) \otimes \alpha_2(x)) \
\multimap \ (\dynex x:\beta. \alpha_1(x)) \otimes \dynex x:\beta.
\alpha_2(x)$
\end{propA}

In fact, each of the above formulas can be given graphical
interpretations that correspond to basic breaches of the DPO
conditions \cite{ToHeICE}.

\subsection{Proof rules} \label{section:rules}

\[
\begin{array}{lll}
  \infer [LId] {[\emptyset]; \Gamma; u::\alpha \vdash u::\alpha  
                     \quad \mbox{with} \ \alpha \ \mbox{atomic} } {}  
      \quad & \quad 

  \infer [UId] {[\emptyset]; \Gamma, x::\alpha; \cdot \vdash x::\alpha 
                     \quad \mbox{with} \ \alpha \ \mbox{closed}  } {} 
\end{array}
\]

\[
\begin{array}{l}
  \infer [\dynex R]
    { [\Sigma_1,\Sigma_2,FV(D)]; \Gamma_1, \Gamma_2; 
      \Delta, n:: \beta \downmapsto D \vdash \lnvareps (D|n). 
      M :: \dynex x: \beta. \alpha 
    }  
    {
    \begin{array}{ll}
      [\Sigma_2]; \Gamma_2, x::\beta; \cdot \vdash N:: 
         \alpha \multimap \alpha 
  \\
    \ [ \Sigma_1] ; \Gamma_1 ; 
        \cdot \vdash D:: \beta 
   &  [\Sigma_1,\Sigma_2]; \Gamma_1, \Gamma_2; \Delta \vdash M ::
        \alpha[D / x] 
    \end{array} }  \\
  \\  
  \infer [\dynex L] { [\Sigma]; \Gamma; 
      \Delta,  u:: \dynex z:
      \beta. \ \alpha \vdash \msf{let} 
      \ \lnvareps
      (z|n). v = u \ \msf{in} \ N:: \gamma }
    { [\Sigma,z]; \Gamma, z::\beta; \Delta, n::\beta \downmapsto z, 
      v::\alpha \vdash N:: \gamma }  
\end{array}
\]

\[
\begin{array}{ll}
\infer [\forall R] { [\Sigma - x ]; \Gamma; 
          \Delta \vdash \lambda x. \ M:: \forall
  x:\beta. \ \alpha } { [\Sigma]; \Gamma, x::\beta; 
      \Delta \vdash M:: \alpha }  
&
  \infer [\multimap R] { [\Sigma]; \Gamma;  
      \Delta \vdash \lnlambda u:\alpha. \ M::
    \alpha \multimap \beta } { [\Sigma]; \Gamma;  
          \Delta, u:: \alpha \vdash M:: \beta
  }  
\end{array}
\]

\[
\begin{array}{l}
  \infer [\forall L]
  { [\Sigma_1,\Sigma_2]; \Gamma;   
       \Delta, u:: \forall x:\beta. \alpha \vdash \letexp{v}{u D}{N}:: 
       \gamma } { [\Sigma_1]; \Gamma; \cdot \vdash D:: \beta    
          &  [\Sigma_2]; \Gamma; \Delta, 
                   v:: \alpha[D/x] \vdash N:: \gamma 
  }
\end{array}
\]

\[
\begin{array}{l}
  \infer [\multimap L] {
\begin{array}{ll}
 [\Sigma_1, \Sigma_2]; \Gamma; 
        \Delta_1, \Delta_2, v::\alpha \multimap \beta 
 \vdash \letexp{u}{v \caret M}{N}:: \gamma
\end{array}} 
        { 
            [\Sigma_1]; \Gamma; \Delta_1 \vdash M::\alpha
        & 
             [\Sigma_2]; \Gamma; \Delta_2, u::\beta \vdash N::\gamma 
        }
\end{array}
\]

\[
\begin{array}{ll}
  \infer [\otimes R] { [\Sigma_1, \Sigma_2]; \Gamma; 
           \Delta_1, \Delta_2 \vdash M \otimes N::
         \alpha \otimes \beta } {  
         [\Sigma_1];  \Gamma;  
          \Delta_1 \vdash M::\alpha 
        & [\Sigma_2]; \Gamma; \Sigma_2;
    \Delta_2 \vdash N::\beta }  
&
  \infer [\otimes L]
  { [\Sigma]; \Gamma; \Delta, w::\alpha \otimes \beta \vdash
    \letexp{u \otimes v}{w}{N}:: \gamma }
  { [\Sigma]; \Gamma; \Delta, u:: \alpha, v:: \beta \vdash N:: \gamma } 
\end{array}
\]

\[
\begin{array}{ll}
  \infer [\mathbf{1}R] { [\emptyset]; \Gamma;  
             \cdot \vdash \msf{nil} :: \mathbf{1} } {}
  &
  \qquad \infer [\mathbf{1}L] { [\Sigma]; \Gamma; \Delta, u::\mathbf{1} \vdash
    \letexp{\msf{nil}}{u}{N}::\alpha } { [\Sigma]; \Gamma; 
         \Delta \vdash N::\alpha }
\end{array}
\]

\[
\begin{array}{ll}
\infer [! R] { [\Sigma]; \Gamma; \cdot \vdash \ !M :: \ !\alpha } 
      { [\Sigma]; \Gamma; \cdot
  \vdash M:: \alpha } 
  & \qquad \infer [! L] { [\Sigma]; \Gamma; 
        \Delta, u::!\alpha \vdash \letexp{!x}{u}{N}::\beta } 
     { [\Sigma]; \Gamma, x::\alpha; \Delta \vdash N::\beta }
\end{array}
\]

\[
\begin{array}{ll}
  \infer [Weak] {\begin{array}{ll}
 [\Sigma]; \Gamma, \Gamma'; \Delta  \vdash  
       \msf{discard}(\Gamma') \ \msf{in} \ N:: \alpha
\end{array}
} 
{ [\Sigma]; \Gamma; \Delta \vdash  N:: \alpha} 
&
\qquad  \infer [Contr] {\begin{array}{ll}
 [\Sigma]; \Gamma, x::\alpha; \Delta  \vdash  
    \letexp{u}{\msf{copy}(x)}{N}::\gamma 
 \end{array}
} 
{ [\Sigma]; \Gamma, x::\alpha; \Delta, u::\alpha \vdash N::\gamma }
 \end{array}
\]

\[
\begin{array}{l}
  \infer [Cut] { [\Sigma, \Sigma']; \Gamma; 
 \Delta, \Delta' \vdash \letexp{u}{N}{M}::\beta } 
    { 
          [\Sigma]; \Gamma; \Delta \vdash N::\alpha
        & 
        [\Sigma']; \Gamma; \Delta', u:: \alpha \vdash M::\beta} 
\end{array}
\]

\[
\begin{array}{l}
\infer [!Cut] {[\Sigma,\Sigma'[FV(D)/x] ]; 
        \Gamma; \Delta[D/x] \vdash \letexp{x}{D}{M}::\beta } 
{ [\Sigma]; \Gamma; \cdot \vdash D::\alpha  
  & [\Sigma']; \Gamma, x:: \alpha; \Delta \vdash M::\beta}
 \end{array}
\]


\section{Graphs in HILL} \label{section:Trans}

It is possible to embed GT systems in HILL, along lines given in
\cite{ToHeICE} --- though there the logic allowed only for variables
as naming terms, making it harder to deal with hierarchical graphs.
Here instead a node can be represented as non-linear term $D::T$ where
$T = !A$ and $A$ is an atomic closed type, for which we can assume no
closed terms are given. This makes it possible to deal with granular
representations in which nodes can be subgraphs.

An edge can be represented as a dependently typed function variable $u::
\forall x_1:T_1, \ldots, x_k:T_k.  E(x_1,\ldots,x_k)$.  An edge component can
be derived as a sequent

$$[\Sigma_1,\ldots,\Sigma_k]; \Gamma; u :: \forall x_1:T_1, \ldots, x_k:T_k. 
E(x_1,\ldots,x_k) \vdash u \ D_1 \ldots \ D_k :: E(D_1,\ldots,D_k) $$

\noindent from the assumptions $[\Sigma_1]; \Gamma; \cdot \vdash
D_1::T_1 \quad \ldots \quad [\Sigma_k]; \Gamma; \cdot \vdash D_k::T_k
$.

The same component with hidden node names can be represented as

$$
\begin{array}{ll} 
  [\Sigma']; \Gamma; n_1::T_1 \downmapsto D_1, \ldots, n_k::T_k \downmapsto D_k, 
  u :: \forall x_1:T_1, \ldots, x_k:T_k.  E(x_1,\ldots,x_k)
  \vdash \\
  \qquad \qquad \varepsilon (D_1|n_1) \ldots (D_k|n_k).  u \ x_1  \ldots \ x_k :: \dynex x_1:T_1, \ldots,
  x_k:T_k. E(x_1,\ldots,x_k) 
\end{array}
$$

\noindent where $\Sigma' =
[\Sigma_1,FV(D_1),\ldots,\Sigma_k,FV(D_k)]$. The empty graph can be
represented as $[\emptyset ]; \Gamma; \cdot \vdash \msf{nil}::\One$.
The parallel composition of two components $\ [\Sigma_1]; \Gamma;
\Delta_1 \vdash G_1::\gamma_1 \ $ and $ \ [\Sigma_2]; \Gamma; \Delta_2
\vdash G_2::\gamma_2$

\noindent can be represented as

$$[\Sigma_1,\Sigma_2]; \Gamma; \Delta_1, \Delta_2 \vdash 
  G_1 \otimes G_2::\gamma_1 \otimes \gamma_2$$

  As a further example, assuming $[\Sigma]; \Gamma; \cdot \vdash D::T$
  an isolated node can be represented as

$$[\Sigma,FV(D)]; \Gamma;n::T \downmapsto D \vdash \varepsilon (D|n). \msf{nil}::
\dynex x:T. \One$$

It is not difficult to see how an encoding of hypergraphs into HILL
can be defined inductively along these lines. Let $G$ be a typed
hypergraph, and let it be closed (i.e. without external nodes). We
can define a graph signature $\langle \Delta^N_G,\Delta^E_G \rangle$,
where $\Delta^N_G$ are the locations that represent the nodes
of $G$, and $\Delta^E_G$ are the linear variables that represent
the edges of $G$. We call \emph{graph formulas} those in the
$\mathbf{1},\otimes,\dynex, \forall, \downmapsto$ fragment
of the logic containing as primitive types only node and edge types,
such that quantification ranges on node types only. We say that a
graph formula $\gamma$ is in normal form whenever $\gamma \ = \ \dynex
(\overline{x:T}). \ \alpha$, where either $\alpha = \mathbf{1}$ or
$\alpha = E_1(\overline{x}_1) \otimes \ldots \otimes
E_k(\overline{x}_k) $, with $\overline{x::T}$ a sequence of typed
variables. The formula is closed if $\overline{x}_{i} \subseteq
\overline{x}$ for each $1 \leq i \leq k$.  $G$ can be represented by a
derivation

$$[FN(\Delta^N_G)]; \Gamma;\Delta^N_G,\Delta^E_G \vdash N_G::\gamma$$

where $\gamma$ is a closed normal graph formula that we call
\emph{representative} of $G$. This encoding can be extended to an
abstract hypergraphs $I \to G$, by representing edges with linear
variables $\Delta^E_G$ and internal nodes with locations $\Delta^N_G$
as before, and by representing interface nodes as free variables that
can be $\lambda$-abstracted. The representative $\gamma$ has then form
$\forall x_1:T_1,\ldots,x_j:T_j.  \gamma'$, where $\gamma'$ is a
normal graph formula, and $x_1:T_1,\ldots,x_j:T_j$ are the open nodes.
This translation generalises that given in \cite{ToHeICE}.

\subsection{Transformation rules} \label{TR}

Graph transformation can be represented by linear inference.  In
particular, a direct transformation $G \Rrel{} H$, where $G,H$ are
closed hypergraphs, can be encoded logically as $\gamma_G \multimap
\gamma_H$, where $\gamma_G,\gamma_H$ are representatives of $G$ and
$H$, respectively.  Let $\pi(p) = \Lambda K. L \Rrel{} R$ be a DPO
transformation rule with discrete interface, i.e. such that $K$ is the
set of the typed nodes that are shared between $L$ and $R$, and such
that none of them is isolated in both $L$ and $R$. Then $p$ can be
represented logically as non-linear term

$$z_p::!\forall x_1:T_1,\ldots,x_k:T_k. \gamma_L \multimap \gamma_R$$

where $\gamma_L,\gamma_R$ are normal graph formulas, representatives of $L$
and $R$ respectively, and $x_1:T_1,\ldots,x_k:T_k$ represent the nodes in $K$.
The ! closure guarantees unrestricted applicability, universally quantified
variables represent the rule interface, and linear implication represents
transformation.


As shown in the double-pushout diagram (section \ref{section:GTS}), the
application of rule $\pi(p)$ determined by morphism $m$ to a closed hypergraph
$G$, resulting in a closed hypergraph $H$, can be represented up to
isomorphism as a derivation of an $H$ representative $\alpha_H = \dynex
\overline{y:T_y}. \beta_H$ from a $G$ representative $\alpha_G = \dynex
\overline{y:T_y}. \beta_G$, based on $z_p$ and on the multiple substitution
$[\overline{z:T_z} \lrel d \overline{x:T_x}]$ of the free variables in
$\gamma_L, \gamma_R$, corresponding to the interface morphism $d$ (not
required to be injective) in the diagram.  A transformation determined by an
application of the rule can be proved correct, up to isomorphism, by the fact
that the following is a derivable rule

$$
\infer [\Rrel{p,m}] { \ [\emptyset]; \Gamma; \forall \overline{x:T_x}.
    \alpha_L \ \multimap \ \alpha_R \vdash \alpha_G \multimap \alpha_H }
{
  \begin{array}{llll} 
    \ [\emptyset]; \Gamma; \cdot \vdash \alpha_G \lequiv \alpha_{G'} & \alpha_{G'} =
    \dynex \overline{y:T_y}. \alpha_L [\overline{z:T_z} \lrel d
    \overline{x:T_x}]
    \otimes \alpha_C \\
    \ [\emptyset]; \Gamma; \cdot \vdash \alpha_H \lequiv \alpha_{H'} & \alpha_{H'} =
    \dynex \overline{y:T_y}. \alpha_R
    [\overline{z:T_z} \lrel d \overline{x:T_x}]
    \otimes \alpha_C
  \end{array}
}
$$

where $\overline{z:T_z} \subseteq \overline{y:T_y}$, as $G$ and $H$
are closed.


\begin{propA} \label{obs:20} The application of a transformation rule
  to a closed graph representative implies linearly a closed graph
  representative that is determined up to graph isomorphism by the
  instantiation of the rule interface variables (morphism $d$). The
  match determined by $d$ (up to isomorphism) satisfies the gluing
  condition on both sides --- with respect to the rule instance
  premise and the initial graph, and with respect to the rule instance
  consequence and the resulting graph --- and therefore satisfies the
  DPO conditions (Proof: since $\dynex$ behaves injectively with
  respect to multiple instantiations, as from Prop. \ref{obs:51}(2.b),
  and satisfies the properties of restriction in Prop. \ref{obs:5}).
\end{propA}

As to reachability, a sequent $\Gamma ; P_1,\ldots,P_k, G_0 \vdash
G_1$, where $\Gamma$ does not contain any rule, can express that graph
$G_1$ is reachable from the initial graph $G_0$ by applying rules $P_1
= \forall \overline{x}_1.  \alpha_1 \multimap \beta_1, \ \ldots, \ P_k
= \forall \overline{x}_k.  \alpha_k \multimap \beta_k$ once each,
abstracting from the application order. A sequent $\Gamma,P_1,\ldots, P_k;
G_0 \vdash G_1$ can express that $G_1$ is reachable from $G_0$ by the
same rules, regardless of whether or how many times they are applied.
The parallel application of rules $\forall \overline{x}_1.  \alpha_1
\multimap \beta_1$, $\forall \overline{x}_2.  \alpha_2 \multimap
\beta_2$ can be represented as application of $\forall \overline{x}_1
\overline{x}_2. (\alpha_1 \multimap \beta_1) \otimes (\alpha_2
\multimap \beta_2)$, as distinct from $\forall \overline{x}_1
\overline{x}_2. \alpha_1 \otimes \alpha_2 \multimap \beta_1 \otimes
\beta_2$.

\subsection{Example}

We give an example of logic derivation that represents the application
of a transformation rule (graphically represented in Fig.  1),
conveniently simplifying the notation, by making appropriate naming
choices.

$$
\infer[\otimes L, \dynex L^*]
{\begin{array}{ll}
\Gamma; n_1\downmapsto x_1,n_2 \downmapsto x_2, 
 A(x_1,x_2), B(x_2), \\
 \dynex y_3, y_4:\alpha_3. C(x_1)
\otimes D(y_3,y_4) \vdash \gamma_H \\
\qquad \qquad \vdots 
\end{array}}
{\infer[\dynex R^*]
{
\begin{array}{ll}
\Gamma^* ; 
n_1\downmapsto x_1,n_2 \downmapsto x_2,n_5\downmapsto x_5,n_6 \downmapsto x_6, \\
 A(x_1,x_2), B(x_2), 
  C(x_1), D(x_5,x_6) \vdash \gamma_H \\
\quad \Gamma^* = \Gamma, x_5,  x_6
\end{array}
}
{
\begin{array}{ll}
\infer[\otimes R^*]
{\begin{array}{ll}
\Gamma^* ; 
 A(x_1,x_2), B(x_2), 
  C(x_1), D(x_5,x_6) \vdash  \\
\qquad C(x_1) \otimes A(x_1,x_2) \otimes D(x_5,x_6) 
  \otimes B(x_2)
\end{array}
}
{ 
\begin{array}{ll}
\Gamma^*;  
 A(x_1,x_2) \vdash A(x_1,x_2) \\
\Gamma^*; B(x_2) \vdash B(x_2) \\
\Gamma^*; C(x_1) \vdash  C(x_1) \\
\Gamma^*; D(x_5,x_6) \vdash D(x_5,x_6) 
\end{array}
}
\\
 \Gamma^* \vdash  x_1 \qquad 
 \Gamma^* \vdash  x_2  \\
 \Gamma^* \vdash  x_5 \qquad
 \Gamma^* \vdash  x_6 
\end{array}
}
}
$$

$$
\infer[\multimap R]{\Gamma'; \delta \vdash \gamma_G 
    \multimap \gamma_H}
{\infer[\dynex L^*]{\Gamma'; \delta, \gamma_G \vdash \gamma_H}
{\begin{array}{ll}
\infer[\forall L]
{
 \Gamma; n_1\downmapsto x_1,n_2 \downmapsto x_2,n_3 \downmapsto
  x_3, C(x_1), A(x_1,x_2), A(x_1,x_3), B(x_2), \delta \vdash \gamma_H }
{\infer[\multimap L]
{\begin{array}{ll}
\Gamma; n_1\downmapsto x_1,n_2 \downmapsto x_2,n_3 \downmapsto
  x_3,  
 C(x_1), A(x_1,x_2), A(x_1,x_3), B(x_2), \\
 (\dynex y_2:\alpha_2.  
C(x_1) \otimes A(x_1,y_2)) \multimap (\dynex y_3, y_4:\alpha_3. C(x_1)
\otimes D(y_3,y_4)) \vdash \gamma_H 
\end{array}
}
{\infer[\dynex R]
{\begin{array}{ll}
\Gamma; n_3 \downmapsto x_3, 
    C(x_1), A(x_1,x_3) \\
\qquad \vdash \dynex y_2:\alpha_2.  
C(x_1) \otimes A(x_1,y_2)
\end{array}
}
{\begin{array}{ll}
\infer[\otimes R]
{\begin{array}{ll}
\Gamma; 
    C(x_1), A(x_1,x_3) \\
             \qquad \vdash  
C(x_1) \otimes A(x_1,x_3) 
\end{array}
}
{
\begin{array}{ll}
\Gamma; 
    A(x_1,x_3) \vdash  
 A(x_1,x_3)\\
\Gamma; 
    C(x_1) \vdash  
C(x_1)
\end{array}
}
\\
\Gamma \vdash x_3
\end{array}
}
&
\begin{array}{ll}
\qquad \qquad \vdots \\
\Gamma; n_1\downmapsto x_1,n_2 \downmapsto x_2, \\
 A(x_1,x_2), B(x_2), \\
 \dynex y_3, y_4:\alpha_3. C(x_1) \\
\otimes D(y_3,y_4) \vdash \gamma_H 
\end{array}
}
} \\
\qquad \qquad \qquad 
   \Gamma = \Gamma', x_1, x_2, x_3
\end{array}
}
}
$$\\

\noindent where graphs $G, H$ and rule $\pi(p)$ be represented as follows

$\gamma_G = \dynex x_1:\alpha_1, x_2:\alpha_2, x_3:\alpha_3.  
C(x_1) \otimes A(x_1,x_2) \otimes A(x_1,x_3) \otimes B(x_2)$

$\delta = \forall y_1:\alpha_1. (\dynex y_2:\alpha_2.  
C(y_1) \otimes A(y_1,y_2)) \multimap (\dynex y_3, y_4:\alpha_3. C(y_1)
\otimes D(y_3,y_4)$

$\gamma_H = \dynex z_1:\alpha_1, z_2:\alpha_2, z_3 z_4:\alpha_3.  
C(z_1) \otimes A(z_1,z_2) \otimes D(z_3,z_4) \otimes B(z_2)$\\

The derivation shows that the graph represented as $\gamma_H$ can be
obtained by a single application to the graph represented as
$\gamma_G$ of the rule represented as $\delta$ . The transformation
can be represented logically as sequent $\Gamma'; \delta \vdash
\gamma_G \multimap \gamma_H$, easily provable by backward application
of the proof rules, as shown. The fact that naming terms here are
variables makes the book-keeping of free nominal variables
straightforward (and annotation unnecessary).

\section{Conclusion and further work}


We have discussed how to represent DPO-GTS in a quantified extension
of ILL, to reason about concurrency and reachability at the abstract
level. We focussed on abstraction from name identity, an aspect that
in hyperedge replacement formulations of GT is often associated with
name restriction \cite{CHARM}. We used an approach that, with respect
to nominal logic, appears comparatively closer to \cite{ScSt04} than
to \cite{pym02} --- though our resource-bound quantifier is
essentially based on existential quantification, and unlike freshness
quantifiers does not seem to be so easily understood in terms of
\emph{for all}.

We have followed the general lines of the encoding presented in
\cite{ToHeICE}, but we have relied on a more expressive logic,
allowing for the use of complex terms as names. With this extension,
it becomes possible to go beyond flat hypergraphs as defined in
section \ref{section:GTS}, and to consider structured ones
\cite{DHP02,BusattoKK05}. Moreover, it should be possible to deal with
transformation rules that are not discrete, i.e. that include edge
components in the interface, by shifting to a representation in which
hyperedges, too, are treated as names. However, if such extensions do
not appear particularly problematic from the point of view of
soundness, they may make completeness results rather more difficult.


\begin{figure}
\centering{
\includegraphics*[scale=0.60]{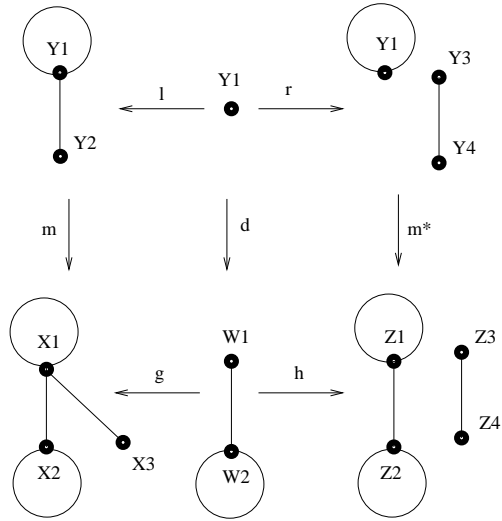}
\caption{Transformation example}\label{Transformation example}}
\end{figure}


\bibliographystyle{../eptcs} 



\begin{thebibliography}{10}
\providecommand{\bibitemstart}[1]{\bibitem{#1}}
\providecommand{\bibitemend}{}
\providecommand{\bibliographystart}{}
\providecommand{\bibliographyend}{}
\providecommand{\url}[1]{\texttt{#1}}
\providecommand{\urlprefix}{Available at }
\providecommand{\bibinfo}[2]{#2}
\bibliographystart

\bibitemstart{abram93}
\bibinfo{author}{S.~Abramsky} (\bibinfo{year}{1993}):
  \emph{\bibinfo{title}{Computational interpretation of linear logic}}.
\newblock {\sl \bibinfo{journal}{Theoretical Computer Science}}
  \bibinfo{volume}{111}.
\bibitemend

\bibitemstart{BeEA93}
\bibinfo{author}{N.~Benton}, \bibinfo{author}{G.~Bierman},
  \bibinfo{author}{V.~de~Paiva} \& \bibinfo{author}{M.~Hyland}
  (\bibinfo{year}{1993}): \emph{\bibinfo{title}{Linear lambda-calculus and
  categorical models revisited}}.
\newblock In: \bibinfo{editor}{E.~B\"{o}rger}, \bibinfo{editor}{G.~J\"{a}ger},
  \bibinfo{editor}{Kleine~H. B\"{u}ning}, \bibinfo{editor}{S.~Martini} \&
  \bibinfo{editor}{M.~Richter}, editors: {\sl \bibinfo{booktitle}{Proceedings
  of the Sixth Workshop on Computer Science Logic}}.
  \bibinfo{publisher}{Springer Verlag}, pp. \bibinfo{pages}{61--84}.
\bibitemend

\bibitemstart{BusattoKK05}
\bibinfo{author}{Giorgio Busatto}, \bibinfo{author}{Hans-J{\"o}rg Kreowski} \&
  \bibinfo{author}{Sabine Kuske} (\bibinfo{year}{2005}):
  \emph{\bibinfo{title}{Abstract hierarchical graph transformation}}.
\newblock {\sl \bibinfo{journal}{Mathematical Structures in Computer Science}}
  \bibinfo{volume}{15}(\bibinfo{number}{4}), pp. \bibinfo{pages}{773--819}.
\bibitemend

\bibitemstart{CePf02}
\bibinfo{author}{I.~Cervesato} \& \bibinfo{author}{F.~Pfenning}
  (\bibinfo{year}{2002}): \emph{\bibinfo{title}{A linear logical framework}}.
\newblock {\sl \bibinfo{journal}{Information and Computation}}
  \bibinfo{volume}{179(1)}, pp. \bibinfo{pages}{19--75}.
\bibitemend

\bibitemstart{CeSc06}
\bibinfo{author}{Iliano Cervesato} \& \bibinfo{author}{Andre Scedrov}
  (\bibinfo{year}{2006}): \emph{\bibinfo{title}{Relating State-Based and
  Process-Based Concurrency through Linear Logic}}.
\newblock {\sl \bibinfo{journal}{Electron. Notes Theor. Comput. Sci.}}
  \bibinfo{volume}{165}, pp. \bibinfo{pages}{145--176}.
\bibitemend

\bibitemstart{Clarke07}
\bibinfo{author}{David Clarke} (\bibinfo{year}{2007}):
  \emph{\bibinfo{title}{Coordination: {R}eo, nets, and logic}}.
\newblock In: {\sl \bibinfo{booktitle}{FMCO 2007}}, {\sl
  \bibinfo{series}{LNCS}} \bibinfo{volume}{5382}. pp.
  \bibinfo{pages}{226--256}.
\bibitemend

\bibitemstart{CHARM}
\bibinfo{author}{Andrea Corradini}, \bibinfo{author}{Ugo Montanari} \&
  \bibinfo{author}{Francesca Rossi} (\bibinfo{year}{1994}):
  \emph{\bibinfo{title}{An abstract machine for concurrent modular systems:
  {C}{H}{A}{R}{M}}}.
\newblock {\sl \bibinfo{journal}{Theoretical Computer Science}}
  \bibinfo{volume}{122}, pp. \bibinfo{pages}{165--200}.
\bibitemend

\bibitemstart{Courcelle97}
\bibinfo{author}{B.~Courcelle} (\bibinfo{year}{1997}):
  \emph{\bibinfo{title}{The expression of graph properties and graph
  transformation in monadic second-order logic}}.
\newblock In: \bibinfo{editor}{G.~Rozenberg}, editor: {\sl
  \bibinfo{booktitle}{Handbook of Graph Grammars and Computing by Graph
  Transformation}}, ~\bibinfo{volume}{1}. \bibinfo{publisher}{World
  Scientific}, pp. \bibinfo{pages}{313--400}.
\bibitemend

\bibitemstart{Dixon06a}
\bibinfo{author}{Lucas Dixon}, \bibinfo{author}{Alan Smaill} \&
  \bibinfo{author}{Alan Bundy} (\bibinfo{year}{2006}):
  \emph{\bibinfo{title}{Planning as Deductive Synthesis in Intuitionistic
  Linear Logic}}.
\newblock \bibinfo{type}{Technical Report}, \bibinfo{institution}{University of
  Edinburgh}.
\bibitemend

\bibitemstart{DoPl08}
\bibinfo{author}{Mike Dodds} \& \bibinfo{author}{Detlef Plump}
  (\bibinfo{year}{2008}): \emph{\bibinfo{title}{From hyperedge relpacement to
  separation logic and back}}.
\newblock In: {\sl \bibinfo{booktitle}{ICGT 2008 --- Doctoral Symposium}}.
\bibitemend

\bibitemstart{DHP02}
\bibinfo{author}{Frank Drewes}, \bibinfo{author}{Berthold Hoffmann} \&
  \bibinfo{author}{Detlef Plump} (\bibinfo{year}{2002}):
  \emph{\bibinfo{title}{Hierarchical graph transformation}}.
\newblock {\sl \bibinfo{journal}{J. Comput. Syst. Sci.}}
  \bibinfo{volume}{64}(\bibinfo{number}{2}), pp. \bibinfo{pages}{249--283}.
\bibitemend

\bibitemstart{EEPT06}
\bibinfo{author}{H.~Ehrig}, \bibinfo{author}{K.~Ehrig},
  \bibinfo{author}{U.~Prange} \& \bibinfo{author}{G.~Taentzer}
  (\bibinfo{year}{2006}): \emph{\bibinfo{title}{Fundamentals of algebraic graph
  transformation}}.
\newblock \bibinfo{publisher}{Springer}.
\bibitemend

\bibitemstart{gabbay04}
\bibinfo{author}{Murdoch~J. Gabbay} \& \bibinfo{author}{J.~Cheney}
  (\bibinfo{year}{2004}): \emph{\bibinfo{title}{A Sequent Calculus for Nominal
  Logic}}.
\newblock In: {\sl \bibinfo{booktitle}{19th Annual IEEE Symposium on Logic in
  Computer Science ({LICS} 2004)}}. pp. \bibinfo{pages}{139--148}.
\bibitemend

\bibitemstart{HirschM99}
\bibinfo{author}{Dan Hirsch} \& \bibinfo{author}{Ugo Montanari}
  (\bibinfo{year}{1999}): \emph{\bibinfo{title}{Consistent transformations for
  software architecture styles of distributed systems}}.
\newblock {\sl \bibinfo{journal}{Electr. Notes Theor. Comput. Sci.}}
  \bibinfo{volume}{28}.
\bibitemend

\bibitemstart{Mil92}
\bibinfo{author}{D.~Miller} (\bibinfo{year}{1992}): \emph{\bibinfo{title}{The
  pi-calculus as a theory in linear logic: preliminary results}}.
\newblock In: {\sl \bibinfo{booktitle}{Workshop on Extensions of Logic
  Programming}}, number \bibinfo{number}{660} in \bibinfo{series}{LNCS}.
  \bibinfo{publisher}{Springer}, pp. \bibinfo{pages}{242--264}.
\bibitemend

\bibitemstart{Pf94}
\bibinfo{author}{Frank Pfenning} (\bibinfo{year}{1994}):
  \emph{\bibinfo{title}{Structural Cut Elimination in Linear Logic}}.
\newblock \bibinfo{type}{Technical Report}, \bibinfo{institution}{Carnagie
  Mellon University}.
\bibitemend

\bibitemstart{Pf02}
\bibinfo{author}{Frank Pfenning} (\bibinfo{year}{2002}):
  \emph{\bibinfo{title}{Linear Logic --- 2002 Draft}}.
\newblock \bibinfo{type}{Technical Report}, \bibinfo{institution}{Carnagie
  Mellon University}.
\bibitemend

\bibitemstart{PiSt93}
\bibinfo{author}{Andrew Pitts} \& \bibinfo{author}{Ian Stark}
  (\bibinfo{year}{1993}): \emph{\bibinfo{title}{Observable Properties of Higher
  Order Functions that Dinamycally Create Local Names, or: What's new?}}
\newblock In: {\sl \bibinfo{booktitle}{MFCS'93}}.
  \bibinfo{publisher}{Springer}, pp. \bibinfo{pages}{122--141}.
\bibitemend

\bibitemstart{Pitts01}
\bibinfo{author}{Andrew~M. Pitts} (\bibinfo{year}{2001}):
  \emph{\bibinfo{title}{Nominal Logic: A First Order Theory of Names and
  Binding}}.
\newblock In: {\sl \bibinfo{booktitle}{TACS '01: Proc. 4th Int. Symp. on
  Theoretical Aspects of Computer Software}}. \bibinfo{publisher}{Springer},
  pp. \bibinfo{pages}{219--242}.
\bibitemend

\bibitemstart{pym02}
\bibinfo{author}{D.~J. Pym} (\bibinfo{year}{2002}): \emph{\bibinfo{title}{The
  semantics and proof-theory of the logics of bunched implications}}.
\newblock Applied Logic Series. \bibinfo{publisher}{Kluwer}.
\bibitemend

\bibitemstart{ScSt04}
\bibinfo{author}{Ulrich Schoepp} \& \bibinfo{author}{Ian Stark}
  (\bibinfo{year}{2004}): \emph{\bibinfo{title}{A Dependent Type Theory with
  Names and Binding}}.
\newblock In: {\sl \bibinfo{booktitle}{Computer Science Logic '04}}.
  \bibinfo{publisher}{Springer}, pp. \bibinfo{pages}{235--249}.
\bibitemend

\bibitemstart{ToHeICE}
\bibinfo{author}{Paolo Torrini} \& \bibinfo{author}{Reiko Heckel}
  (\bibinfo{year}{2009}): \emph{\bibinfo{title}{Towards an embedding of Graph
  Transformation in Intuitionistic Linear Logic}}.
\newblock {\sl \bibinfo{journal}{CoRR}} \bibinfo{volume}{abs/0911.5525}.
\bibitemend

\bibitemstart{ProofTh00}
\bibinfo{author}{A.~S. Troelstra} \& \bibinfo{author}{H.~Schwitchtenberg}
  (\bibinfo{year}{2000}): \emph{\bibinfo{title}{Basic Proof Theory}}.
\newblock \bibinfo{publisher}{Cambridge University Press}.
\bibitemend

\bibliographyend
\end{thebibliography}

\end{document}